\documentclass[a4paper,11pt]{article}
\pdfoutput=1 

\usepackage{jinstpub} 

\usepackage{xfrac}
\usepackage{caption}

\title{Test of SensL SiPM coated with NOL-1 wavelength shifter in liquid xenon}

\author[a,b]{D.\,Yu.\,Akimov,}
\author[a,b,1]{V.\,A.\,Belov,\note{Corresponding author.}}
\author[c]{O.\,V.\,Borshchev,}
\author[a,b]{A.\,A.\,Burenkov,}
\author[a]{Yu.\,L.\,Grishkin,}
\author[a]{A.\,K.\,Karelin,}
\author[a]{A.\,V.\,Kuchenkov,}
\author[a]{A.\,N.\,Martemiyanov,}
\author[c]{S.\,A.\,Ponomarenko,}
\author[a,b,d]{G.\,E.\,Simakov,}
\author[a]{V.\,N.\,Stekhanov,}
\author[c]{N.\,M.\,Surin,}
\author[e]{V.\,S.\,Timoshin,}
\author[a]{O.\,Ya.\,Zeldovich}

\affiliation[a]{Institute for Theoretical and Experimental Physics named by A.\,I.\,Alikhanov of 
National Research Center ``Kurchatov Institute'', 25 Bolshaya Cheremushkinskaya 
st., 117218 Moscow, Russian Federation}
\affiliation[b]{National Nuclear Research University ``MEPhI'', 31 Kashirskoe 
sh., 115409 Moscow, Russian Federation}
\affiliation[c]{Enikolopov Institute of Synthetic Polymer Materials, Russian 
Academy of Science, 70 Profsoyuznaya st., 117393, Moscow, Russian Federation}
\affiliation[d]{Moscow Institute of Physics and Technology, 9 Institutskij per., 
141700, Moscow reg. Russian Federation}
\affiliation[e]{Azimuth Photonics, 11 Khavskaya st., 115162, Moscow, Russian 
Federation}

\emailAdd{belov@itep.ru}

\abstract{A SensL MicroFC-SMT-60035 6$\times$6 mm$^2$ silicon photo-multiplier coated 
with a NOL-1 wavelength shifter 
have been tested in the liquid xenon to detect the 175-nm scintillation light. 
For comparison, a Hamamatsu vacuum ultraviolet sensitive MPPC VUV3 3$\times$3 mm$^2$ 
was tested under the same conditions.
The photodetection efficiency of $13.1 \pm 2.5$\% and 
$6.0 \pm 1.0$\%, correspondingly, is obtained.}


\begin{document}
\maketitle
\flushbottom

\section{Introduction}
\label{sec:intro}

A silicon photomultiplier (SiPM, MPPC, etc.) technology is very attractive for 
the use in low-background experiments in replacement for photomultiplier tubes 
(PMTs) due to the potentially very low level of radioactivity of semiconductor 
materials. Silicon photomultipliers (here and after, we refer to them as SiPM) 
have a gain of about $10^6$, comparable to that of PMT, have the low operating 
voltage and power consumption. The devices have linear response to the light 
intensity when the number of detected photons is not very high in compare to 
the number of SiPM cells. SiPM is not sensitive to electric or magnetic fields. 
Manufacturing is based on the same technology as for many other devices in 
semiconductor industry, that opens the possibility for inexpensive mass 
production. SiPMs have already replaced regular PMTs in a variety of cases, 
including LHCb SciFi upgrade~\cite{i1}, CMS HCal upgrade~\cite{i2} and 
MEG~II~\cite{i3}. Several collaborations like GERDA, NEXT, nEXO are planning 
to use VUV sensitive SiPMs in their detectors.

The first test of SiPM in liquid xenon was performed in 2005~\cite{r1}. 
Unfortunately, SiPMs don't suit well for noble gas based detectors because the 
emitted light of noble gases has wavelengths in a VUV region (below 200~nm). 
The SiPM microstructure is usually not optimized for such wavelengths; 
moreover, protection layers are usually not transparent in this region. Several 
manufacturers have embarked in an extensive development program in order to 
achieve the reasonable sensitivity to the VUV light. Specially dedicated 
devices (e.g. MPPC) have been developed and they are already on practical stage 
for the use in a liquid xenon (emission wavelength is 175~nm) environment in 
MEG experiment. The alternative way to solve this problem is to use a wavelength 
shifter (WLS).

The aim of this work is to demonstrate experimentally the performance of SensL 
SiPM with a WLS to detect the liquid xenon scintillation light and to 
compare its characteristics with those of the VUV-sensitive Hamamatsu Photonics 
VUV3 MPPC device. It was obtained in our previous studies that the 
photodetection efficiency (PDE) of WLS + SiPM system can reach up to $\approx$ 
50\% of the SiPM PDE in the blue region~\cite{l6}. An important feature of our 
study is that the photodetectors are tested directly in the liquid xenon environment.

\section{Wavelength shifter}
\label{sec:wls}

\begin{figure}[htbp]
\centering
\includegraphics[width=0.8\textwidth]{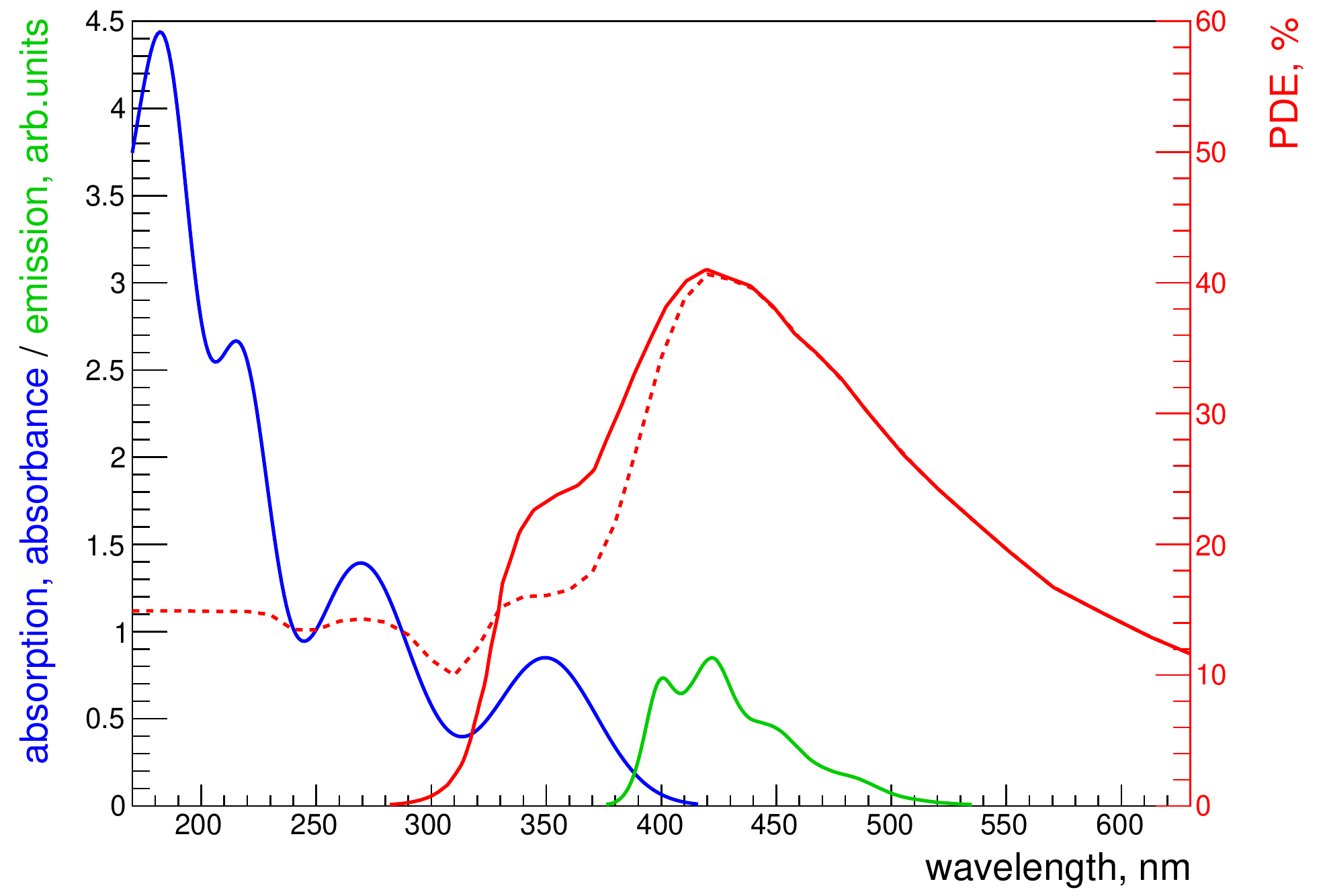}
\caption{\label{fig:nol} Absorption (blue) and emission (green) spectra of 
NOL-1, as well as spectral dependence of SiPM PDE without (red solid) and with 
(red dashed) 200~nm layer of NOL-1.}
\end{figure}

A nanostructured organosilicon luminophore (NOL) is a new type of 
WLS which combines an absorber and an emitter in one molecule. It has 
the very high energy transfer efficiency of the electronic excitation from the 
outer fragments of a molecule absorbing the light at the short wavelengths, to 
the inner fragment having the high photoluminescence quantum yield (QY) at 
longer wavelengths~\cite{l4,l5}. This type of luminophores are of interest for 
various technical applications~\cite{l6,l7}. The NOL-1 type was chosen for 
shifting of the VUV light emission of liquid xenon (175~nm) to the region of 
maximal PDE of SiPM (420~nm). The WLS can be deposited as a thin 
film layer directly on the front surface of SiPM over the protective optically 
transparent compound. Spectral properties of NOL-1 are suited very well for 
this task~\cite{l8}. Absorption spectrum of the transparent 200-nm NOL-1 film 
is plotted in figure~\ref{fig:nol} by blue curve in terms of absorbance 
$A(\lambda)$, where $\lambda$ is a wavelength of light. 
It consists of four absorption bands with maxima at the following wavelengths: 
182~nm, 216~nm, 270~nm and 350~nm. The first three of them correspond to 
excitation of the outer fragments. The absorption band with a maximum at 350~nm 
corresponds to excitation of the central fragment. The emission spectrum 
$P(\lambda)$ of NOL-1 is shown in figure~\ref{fig:nol} by the green curve. It 
corresponds 
to de-excitation of the central fragment and doesn't depend on the excitation 
wavelength. This spectrum has two maxima: at 400 and 422~nm. The effective 
photodetection efficiency $PDE^*(\lambda)$, defined as an overall 
photodetection efficiency of the WLS + SiPM system, is express according to the 
formula:
\begin{equation*}
PDE^*(\lambda) = 
G \cdot \left(1-\frac{1}{10^{A(\lambda)}}\right) \cdot QY(\lambda) \cdot 
\int_{\lambda_{min}}^{\lambda_{max}} P(\lambda') \cdot PDE(\lambda') \cdot d\lambda' + 
\frac{1}{10^{A(\lambda)}} \cdot PDE(\lambda),
\end{equation*}
where $PDE(\lambda)$ is photodetection efficiency of the SensL SiPM (shown in 
figure~\ref{fig:nol} by red), $\lambda_{min}$ and $\lambda_{max}$ are the 
boundaries of the interval where $P(\lambda)>0$. The factor $G=$~0.57--0.61 
describes the amount of emitted light that reach SiPM face. 
It is bigger than a naive assumption value of 0.5 because we also take 
into account reflection of the emitted light from surface of WLS layer 
that comes from the difference in the refractive indices of materials. The 
$QY(\lambda)$ value is approximately equal to 73\% at $\lambda < 300$~nm that 
corresponds to absorption of light by the outer fragments of molecule and equal 
to 80\% for $\lambda$ around 350~nm that corresponds to absorption of light by 
the inner fragment of molecule. The effective photodetection efficiency 
$PDE^*(\lambda)$ calculated with the use of formula above for the 200-nm NOL-1 
layer deposited on the SensL SiPM is shown in figure~\ref{fig:nol} by the red 
dashed curve.

The choice of the 200-nm thickness of NOL-1 layer is based on our previous 
studies~\cite{l6}. The WLS was deposited directly on the front 
surface of SiPM over the protective compound in the form of solution in 
toluene. Then it was dried up forming a solid amorphous film. Such a film has 
higher adhesion to the surface of the protective compound than that of a 
polycrystalline film formed by other frequently used WLSs 
p-terphenyl and tetraphenyl butadiene. This, along with a 7 times larger 
molar mass, makes NOL-1 coating much more steady in liquid xenon, resulting in 
significant reduce of solubility in liquid xenon, compared to p-terphenyl. That 
could make it safe from the point of view of production of a volume distributed 
effects on scintillation light, even when it is in direct contact with liquid 
xenon. 
The absorbance at 175~nm for the WLS layer of such a thickness is higher 
than 4. This means that the layer absorbs 99.99\% of the liquid xenon emission 
light. The absorbance in the region of overlapping of the absorption 
(blue) and the emission (green) spectra is less than $\sim 0.4$. This results 
in reabsorption losses of below 6\%. One can see that the $PDE(\lambda)$ is 
almost zero at the wavelengths shorter than 300~nm. However, depositing of the 
200-nm layer of NOL-1 on SiPM provides $PDE^*$ equal to $\sim$13--15\% at 
these wavelengths. The expected value of $PDE^*$ for 175~nm is $15 \pm 2$\%.

\section{Experimental setup}
\label{sec:setup}

\newcommand{\inchsign}{^{\prime\prime}}

\begin{figure}[htbp]
\centering
\includegraphics[width=0.8\textwidth]{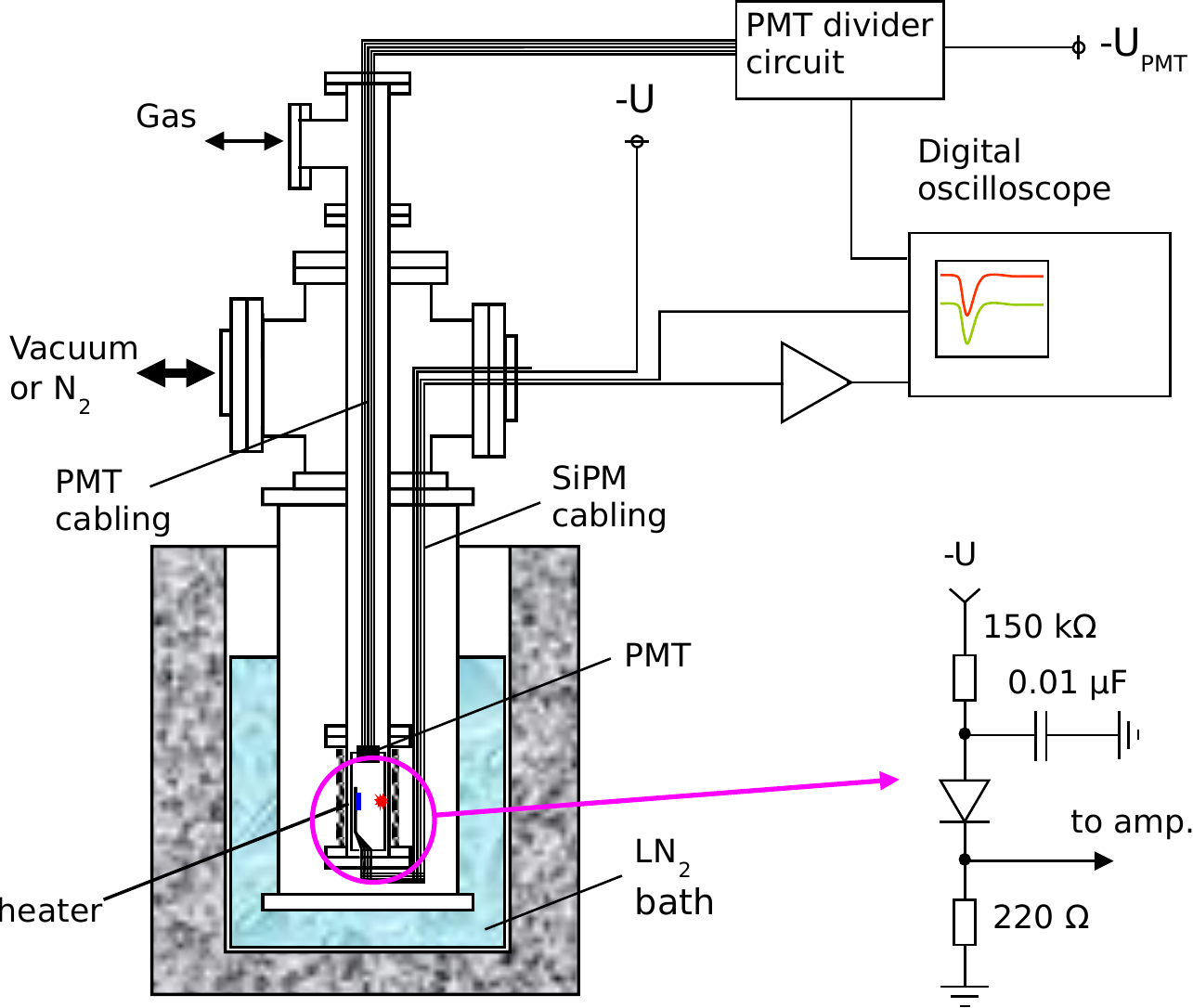}
\caption{\label{fig:schem} Schematic view of test chamber, SiPM electrical 
circuit and data readout.}
\end{figure}

\begin{figure}[htbp]
\parbox[c]{0.39\textwidth}{
\centering
\includegraphics[width=0.30\textwidth]{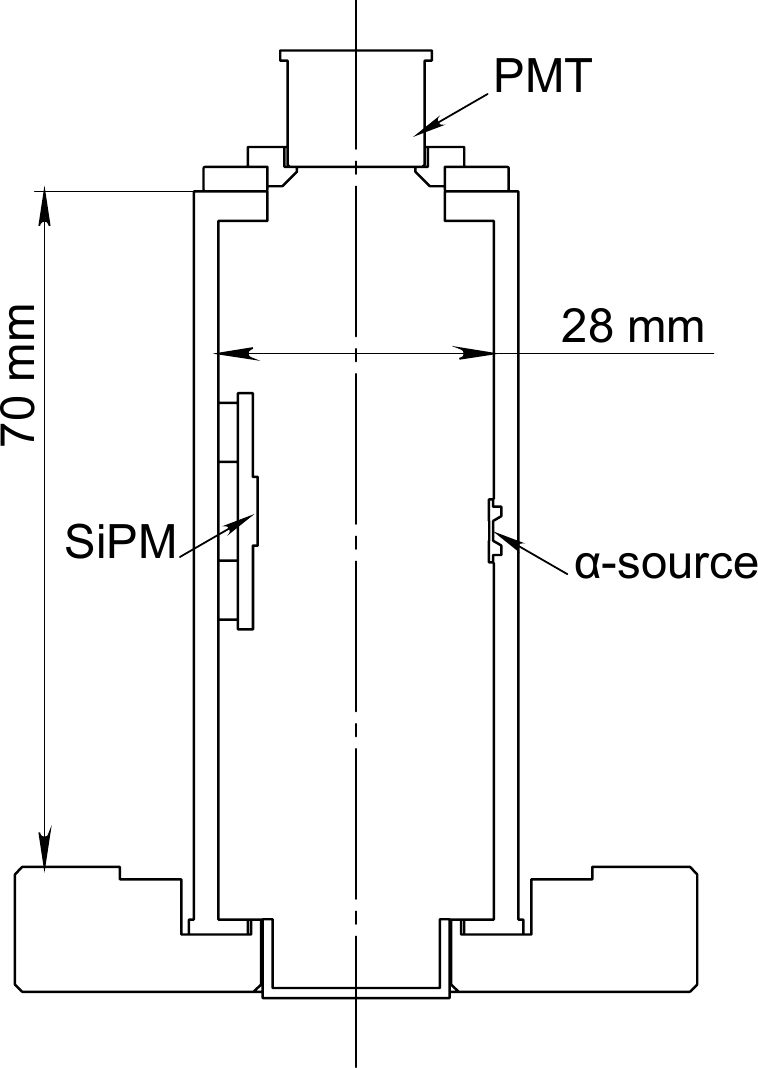}
\caption{\label{fig:assm} Layout of the SiPMs, $\alpha$-source and PMT on the support.}}
\hfill
\parbox[c]{0.59\textwidth}{
\captionof{table}{\label{tab:t1} Individual parameters of studied SiPM's.}
\centering
\begin{tabular}{|l|l|l|}
\hline
 & SensL & Hamamatsu \\
\hline
SiPM model & MicroFC- & VUV3-50um \\
 & 60035-SMT & \\
Size & 6$\times$6~mm$^2$ & 3$\times$3~mm$^2$ \\
Cell size & 35~$\mu$m & 50~$\mu$m \\
Breakdown voltage & 24.5~V & 45.8~V \\
Overvoltage & 5~V & 4~V \\
Solid angle, $\Omega$ & 0.0732~sr & 0.0156~sr \\
Crosstalk, $\alpha_{cr}$ & $18\pm10$\% & $4\pm3$\% \\
\hline
\end{tabular}
}
\end{figure}

The experimental setup used for measurements is shown schematically in figure~\ref{fig:schem}.
The chamber is assembled from standard CF vacuum pieces. The test cell which 
contains SiPM and PMT is made of $2\sfrac{3}{4}\inchsign$ CF nipple (35 mm 
inner diameter) with a feedthrough flange on the bottom. The cell in installed 
in a bigger $8\inchsign$ CF nipple which can be evacuated or filled with 
gaseous N$_2$ for thermo conductivity. A liquid N$_2$ bath is used for cooling the 
chamber. A resistive wire heater wound around the $2\sfrac{3}{4}\inchsign$ CF 
nipple servers for maintaining the required temperature inside the test cell. 
The design of the inner part of the test cell is shown in figure~\ref{fig:assm}.

Inside the cell, there is a construction made of a stainless steel (ss) and 
standing on four ss legs. This construction can hold several SiPMs on the legs 
and a small PMT on the top. An $^{241}$Am alpha-source, which 
produces scintillation light in the liquid xenon, is placed on the leg opposite 
to the SiPM. The SiPM and its electrical circuit are mounted on a ceramic 
plate. A Hamamatsu R7400-06 PMT having a synthetic silica window with bialkali 
photocathode and QE $\sim$15\% for the xenon light was used for triggering. The 
test chamber was pumped out by a zeolite and then by a titanium discharge pump 
down to $\sim 10^{-5}$~torr without being heated. Heating of the cell was not 
allowed because of the photodetector installed inside. Xenon gas had undergone 
purification procedure with a Mykrolis Megaline purifier before filling the 
chamber.

Voltage power supply as well as signal readout was provided with the use of 
coaxial cables and vacuum feedthroughs. The SiPM signals were additionally 
amplified with a gain of 5 by custom made fast amplifiers built on the basis of 
OPA656N OpAmp. For SensL SiPM, the amplifier circuit was located on the back 
side of the ceramic plate; for Hamamtsu SiPM, outside of the cryostat. Signals 
from the SiPMs and PMT were recorded with the Tektronix TDS5034 digital 
oscilloscope at a digitization frequency of 125~MHz. The scope was triggered by 
SiPM in anticoincidence with PMT for noise runs and in coincidence with PMT for 
alpha runs. The threshold in SiPM channel was set low enough to register single 
cell signals. Signal waveforms of a total length of 10~$\mu$s, with trigger 
located in a middle, were recorded for later analysis. 

Two photodetectors were consequently tested: the non-VUV-sensitive SensL MicroFC-60035-SMT 
SiPM coated with the 200-nm NOL-1 WLS layer and the VUV-sensitive 
Hamamatsu Photonics VUV3 MPPC for comparison. 
Characteristics of these photodetectors can be found in table~\ref{tab:t1}. 
The overvoltage for Hamamatsu device was set according to the manufacturer 
recommendations. For SensL SiPM, the overvoltage was selected as a balance 
between gain and noise characteristics to optimize the alpha-peak resolution. 
For each SiPM, three sets of measurements were performed. These include a noise 
run in vacuum plus noise and source runs the liquid xenon environment. All these 
measurements were performed at the temperature of $-100\,^{\circ}\mathrm{C}$.

\section{Data analysis}
\label{sec:anal}

Recorded waveforms of registered events were processed with a dedicated 
software on the event-by-event basis. Signal pulses above the software 
threshold were selected. The software threshold was set depending on noise 
level and gain for each SiPM. For each pulse, a sequence of approximately 100
samples having the amplitude above the threshold defined the pulse 
boundaries and was used to calculate pulse amplitude, area and width. It was 
then baseline-corrected by subtracting the average of the previous 100 samples 
of waveform. 

\begin{figure}[bp]
\centering
\includegraphics[width=.45\textwidth]{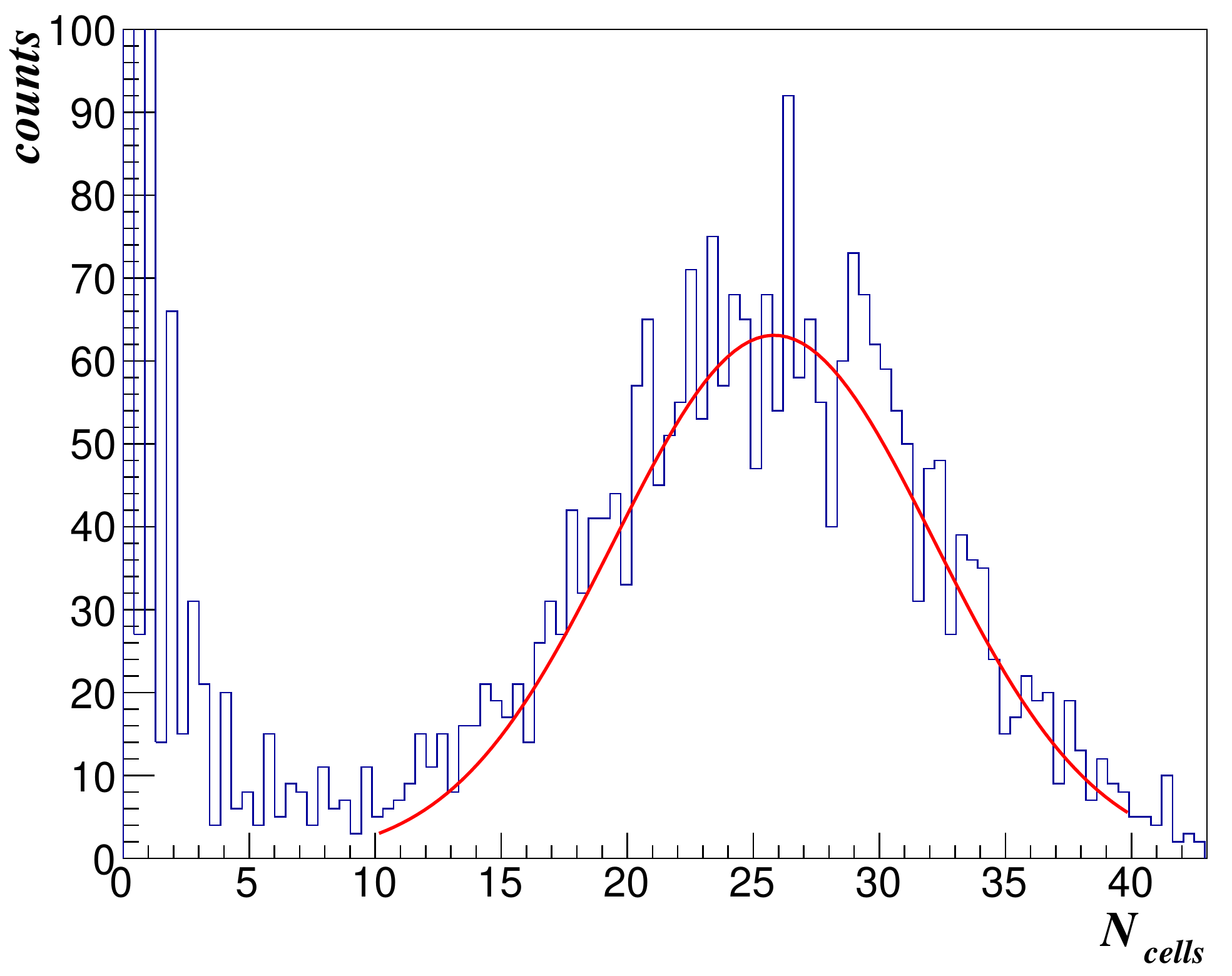}
\hspace{5mm}
\includegraphics[width=.45\textwidth]{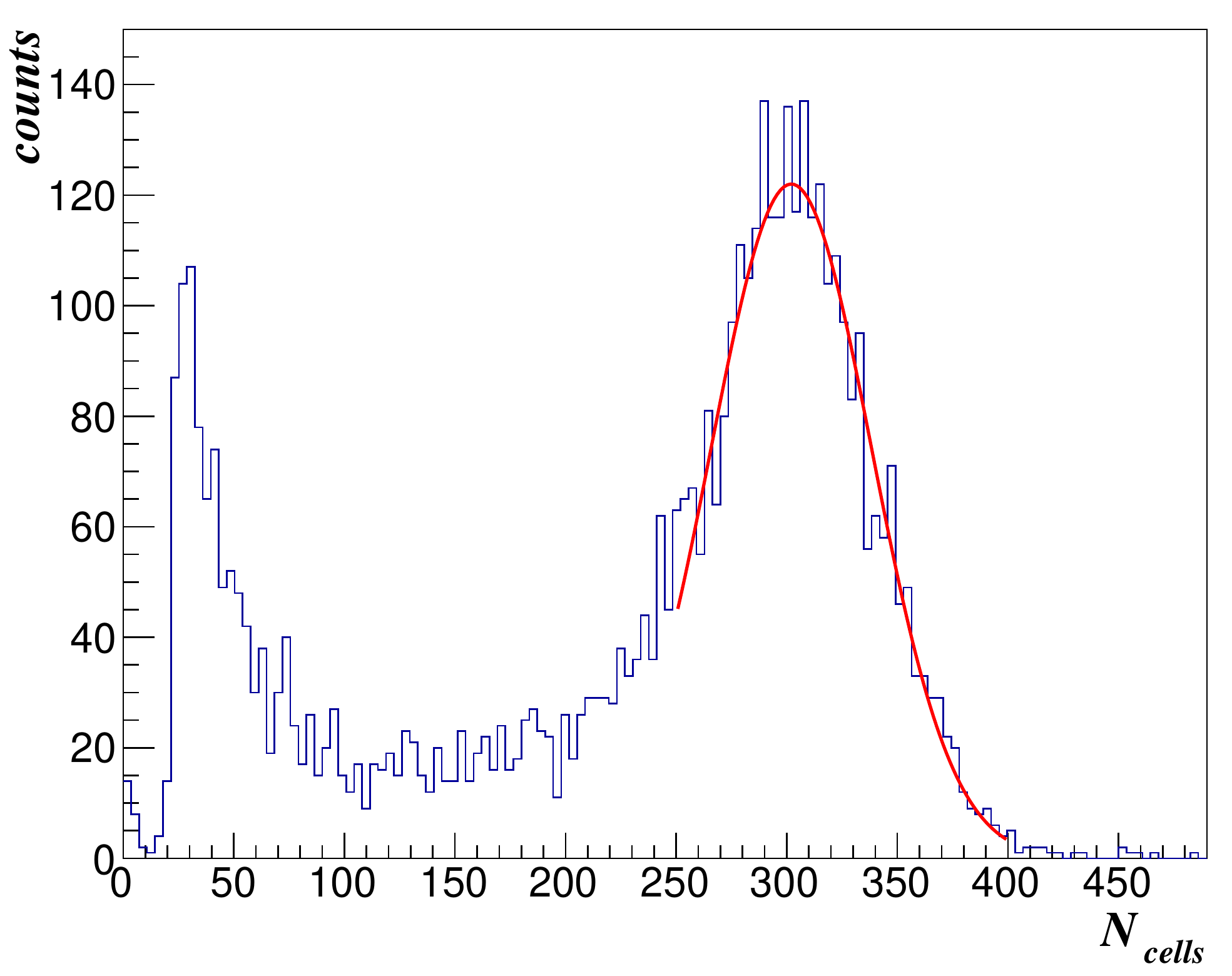}
\caption{\label{fig:data} Area distributions for the signals from Hamamatsu (on 
left) and SensL (on right) SiPMs.}
\end{figure}

Calibration of SiPMs was made by dedicated noise runs in anti-coincidence with PMT 
signal. On the resulting area spectrum, the distance between the first and the 
second peak corresponds to a single cell signal area, and the ratio between the 
second and the first peaks served as a crosstalk estimation. We didn't carry 
out a deep study of afterpulses, but within our time window we did not observe 
any. 

The main datasets with alpha-peak were collected with the use of coincidence with 
PMT. For production of VUV light an $^{241}$Am radioactive source was used. It emits
alpha-particles with a compact set of energies, that results in an average 
energy $E_{\alpha} = 5.48$~MeV. The source was made by shallow ion 
implantation to a substrate i.e.\ with an open surface. Thus, the source emitted 
the alpha-particles almost without energy losses in source material. In liquid 
xenon, the alpha-particles of such energy have the stopping range of 
approximately 40~$\mu$m. Therefore, the radioactive source served as a point-like 
scintillation light source. We used the average energy for scintillation photon 
production by alpha-particles $w = 16.3\pm0.3$~eV~\cite{l9} to convert the 
deposited energy to the 
number of emitted photons. The test chamber and the assembly were made of a 
stainless steel. The chamber walls were unpolished. The reflection coefficient 
for the VUV xenon light (wavelength equals 175~nm) for such a material is known 
to be below 10\%. Thus, we consider a non-direct portion of the light to be 
small enough in our geometry. Consequently, we may estimate the number of 
photons that reach the SiPM surface $N_{ph}$ exclusively by the use of the 
solid angle $\Omega$ spanned from the point of the source location to the SiPM 
photosensitive surface: $N_{ph} = \frac{E_\alpha}{w} \cdot 
\frac{\Omega}{4\pi}$. The resulting SiPM efficiency was calculated as $PDE = 
\frac{\displaystyle{N_{cells}}}{\displaystyle{N_{ph}(1+\alpha_{cr})}}$, where 
$N_{cells}$ is the alpha-peak 
position in terms of number of fired cells, and $\alpha_{cr}$ is estimated 
cross-talk.

The energy spectra measured for alpha-particles in the liquid xenon are shown 
in figure~\ref{fig:data}. From the alpha-peak positions, the PDE values were 
calculated to be 
$6.0\pm1.0$\% and $13.1\pm2.5$\% for Hamamatsu and SensL SiPM, correspondingly. 
The estimated relative error of these PDE values (19\%) is comprised of the 
uncertainty of light collection efficiency (16\%; includes both the error of 
solid angle and unaccounted reflections), error on the crosstalk value (9\%), 
the alpha-particle energy and energy per photon uncertainty ($\sim$5\%).

We couldn't find publications of measurements of Hamamatsu VUV3 in liquid xenon.
Recently, several measurements were performed in liquid xenon, but using different 
device of similar type (Hamamatsu 10943-3186(X) Type A), and showed various 
results. MEG~II experiment presented PDE value to be over 15\% for selected 
samples~\cite{i3}. In another test with this device for DARWIN 
experiment~\cite{e3} author didn't calculate PDE value explicitely, 
but used the same method of measurements. For a 12$\times$12~mm$^2$ 
sample he obtained $N_{cells} \sim$ 300 for alpha-peak with $^{241}$Am source 
installed in 20~mm apart from the SiPM. Given 15 times bigger SiPM surface area, 
one can see that a factor of 12 difference is reasonably compatible with our result.
There is a separate result for the VUV3 measurements in liquid argon, where 
a PDE value of 8\%~\cite{e1} for 128~nm light was demonstrated. Using wavelength 
dependance presented by the manufacturer~\cite{e2}, one can scale this to the 
wavelength of xenon light (175~nm) and can obtain the value 6.8\%. This is also in 
agreement with out result.

\section{Summary}
\label{sec:sum}

We studied the applicability of silicon photomultipliers (SiPM) for the VUV 
light detection in liquid xenon experiments. We have demonstrated that 
the SensL MicroFC-SMT-60035 SiPM coated with the NOL-1 WLS shows 
PDE $13.1\pm2.5$\% for 175~nm light in liquid xenon. 
The VUV sensitive Hamamatsu Photonics VUV3 MPPC device demonstarated PDE 
$6.0\pm1.0$\% under the same conditions. The former result indicates that 
regular commercial SiPMs coated with the NOL-1 WLS can be used 
in liquid xenon detectors as a replacement of PMTs. Such SiPMs are much better 
understood and have a fine-tuned manufacturing process. We developed an easy 
way of NOL-1 thin film deposition on a SiPM surface directly over the 
protective cover. The experimentally measured PDE value for the SensL SiPM 
coated with the NOL-1 is in a good agreement with the expected one.

\acknowledgments

We are very grateful to ``YE International'' (``Hamamatsu Photonics'' 
distributor) and to ``Azimuth Photonics'' (``SensL'' distributor) for supplying 
us with the samples of photodetectors.
This study was supported by RFBR, projects no. 14-02-00675-a, 
14-22-03028-ofi\_m and 15-02-06874-a.


\end{document}